\documentclass[12pt]{article}
\pdfoutput=1

\usepackage{amsmath,amssymb,graphicx}
\usepackage{cite,slashed}
\usepackage[svgnames]{xcolor}
\usepackage{url}
\def\s{\sqrt{\hat{s}}}
\def\kg{\kappa_g}

\def\SM{\text{SM}}
\def\NP{\text{NP}}
\def\A{\mathcal{A}}

\begin{document}

\begin{flushright}
LYCEN 2014-08
\end{flushright}
\vskip.5cm

\begin{center}
{\LARGE \bf Higgs couplings: disentangling New Physics with off-shell measurements}\vskip.1cm
\end{center}
\vskip5mm

\begin{center}
{\bf {Giacomo Cacciapaglia}, {Aldo Deandrea\footnote{also Institut Universitaire de France, 103 boulevard Saint-Michel, 
75005 Paris, France}}, {Guillaume Drieu La Rochelle} and {Jean-Baptiste Flament}}
\end{center}
\vskip 8pt

\begin{center}
{\it
Universit\'e de Lyon, F-69622 Lyon, France; Universit\'e Lyon 1, Villeurbanne;\\
CNRS/IN2P3, UMR5822, Institut de Physique Nucl\'eaire de Lyon\\
F-69622 Villeurbanne Cedex, France} \\
\vspace*{.5cm}
{\tt  g.cacciapaglia@ipnl.in2p3.fr, deandrea@ipnl.in2p3.fr, g.drieu-la-rochelle@ipnl.in2p3.fr, j-b.flament@ipnl.in2p3.fr}
\end{center}\vskip5mm

\begin{abstract}
\vskip 3pt
\noindent
{\small }
After the discovery of a scalar resonance, resembling the Higgs boson, its couplings have been extensively studied via the 
measurement of various production and decay channels on the invariant mass peak. Recently, it has been suggested the 
possibility to use off-shell measurements: in particular, CMS has published results based on the high-invariant mass 
cross section of the process $gg \to ZZ$, which contains the contribution of the Higgs. While this measurement has been interpreted 
as a constraint on the Higgs width after very specific assumptions are taken on the Higgs couplings, in this letter we show that a much 
more model-independent interpretation is possible.
\end{abstract}

Since the discovery of a scalar boson at the LHC by the ATLAS and CMS collaborations, much effort has been devoted to the study of
its properties. So far, most of the information has been obtained by extracting its couplings from the cross-section measurements in 
various production/decay modes on the resonance peak \cite{atlas_higgs,cms_higgs}. Interpreting those results in a Beyond the 
Standard Model (BSM) context can be easily done in terms of a handful of parameters encoding the modification of the couplings to 
standard particles and, among the many proposals, one has been chosen as an official recommendation by experimentalists and 
theorists together \cite{lhcxswg_reco}. 
Recently, a novel kind of measurement has been put forward, where Higgs couplings are extracted from the cross-section integrated 
away from the resonance peak \cite{passarino,melnikov,campbell}, and the first results from CMS on the $H \to Z Z \to 4$ leptons have 
been published \cite{Khachatryan:2014iha}. This new class of measurements is most welcome, since it carries information 
on the Higgs couplings at a different mass scale than its mass shell, and such a dependence on the partonic centre of mass energy $\s$ is paramount 
to distinguishing BSM effects. Although the measurement \cite{Khachatryan:2014iha} was initially proposed as a mean of determining the Higgs 
width, it was quickly recognised in \cite{englert} that such an extraction was too model-dependent to be meaningful. In this short letter, 
we want to point out that the measurement can be primarily interpreted as a bound on the coupling of the Higgs to tops and on the loop contribution of new heavy states to the coupling to gluons, in the limit of 
heavy new physics coloured states. The direct sensitivity to the new loop is novel, as it is not possible to disentangle it from on-peak measurements.\\

In the experimental analysis \cite{Khachatryan:2014iha} the signal region is defined in order to enhance the rate of events coming from the $gg \to ZZ$ process, which consists of a box diagram plus the s-channel Higgs, over the $q\bar{q} \to ZZ$ background. The Higgs contribution consists mainly of a gluon fusion production followed by the decay to two $Z$ bosons, which interferes with the box diagram.
Off-shell, the cross section does not depend explicitly on the Higgs width: however, one can always compensate for a different value of the width in the peak region by rescaling appropriately the couplings of the Higgs to tops and $ZZ$, thus it is the rescaling of the couplings that affects the off-shell cross section.
The interpretation in terms of width, therefore, only works under very specific assumptions: a large contribution to the total width coming from new decay channels, and the absence of new particles running in the Higgs to gluons loop.
The latter point, already properly stressed in \cite{englert}, arises from the fact that at a partonic centre of mass energy away from the Higgs mass peak, 
the loops participating in the production (and decay) of the Higgs may be resolved if the mass of the exchanged particle is light. This is 
the case for Standard Model (SM) loops. To be specific, the measurement in \cite{Khachatryan:2014iha} is dominated 
by the gluon fusion production, which is mainly mediated by a top loop in the SM. This loop form factor does depend on the scale and is 
significantly different at the Higgs mass shell $\s=125.5$ GeV and in an off-peak region $\s>350$ GeV above the top pair threshold. If the 
particle running in the loop were much heavier, say around 1 TeV, the scale dependence would be quite different. In BSM scenarios, one 
can typically expect both modifications of the tree level couplings to the top and new heavy coloured states running in the loop. Due to 
direct searches at the LHC, the mass of the new states is bound to be quite heavy, thus the approximation of heavy new states is 
justified. For an on-shell Higgs, a cancellation between the two effects may and often does occur \footnote{A famous example is the case 
of a heavy vector-like quark mixing to the top via Yukawa-like interactions, as in models of Composite Higgs with top partners.}, so that 
the presence of New Physics may not appear in on-peak measurement. Then it is clear that only measurements at different $\s$ can 
distinguish the two effects since the top and the New Physics contribution have a different scale dependence. In this perspective, the 
parametrisation of the Higgs couplings we proposed in \cite{llodra,flament} is very handy, as it suggests to separate into two independent 
parameters the tree level modification of the top coupling and the effect  of any loop of New Physics states.

The two parameters, relevant for our discussion, are defined as \footnote{In the usually recommended parametrisation \cite{lhcxswg_reco}, 
$\kappa_t$ is defined in the same way, while the coupling to the gluons is defined as $$ \kg=\sqrt{\frac{\Gamma_{h\to gg}}{\Gamma_{h\to gg}^\SM}}\,.$$ On-shell, the parameters in the two sets are simply related: $\kg = \kappa_t + \kappa_{gg}$, while this degeneracy is 
removed off-shell. The difference between the two choices for the coupling to the gluons is crucial for the study of BSM effects from off-shell measurements.}
\begin{equation}
\kappa_t = \frac{g_{h\bar{t} t}}{g_{h \bar{t} t}^{\rm SM}} \qquad \mbox{and} \qquad \kappa_{gg} =\frac{\A_{h\to gg}^\NP}{\A_{h\to gg}^{t,\SM}},
\end{equation}
where $\A_{h\to gg}$ stands for the on-shell amplitude of the process, $^\NP$ stands for the contribution from new particles while $^{t,\SM}$ is the contribution from the top loop with SM coupling to the Higgs. 
The contribution of new physics in the loop is normalised to the top loop following the expectation that the main contribution comes from new particles associated with the top mass generation. The two parameters enter in the calculation of the cross section in two different combinations on-peak and off-shell, so that the two effects can be clearly separated by using this parametrisation.

\begin{figure}[ht]
 \begin{center}
\begin{tabular}{cc}
  \includegraphics[height=6cm]{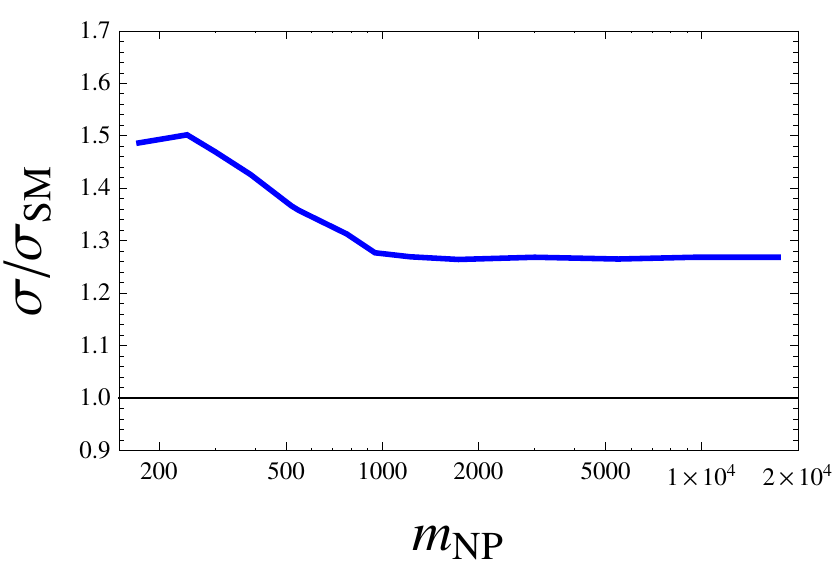}&
  \includegraphics[height=6cm]{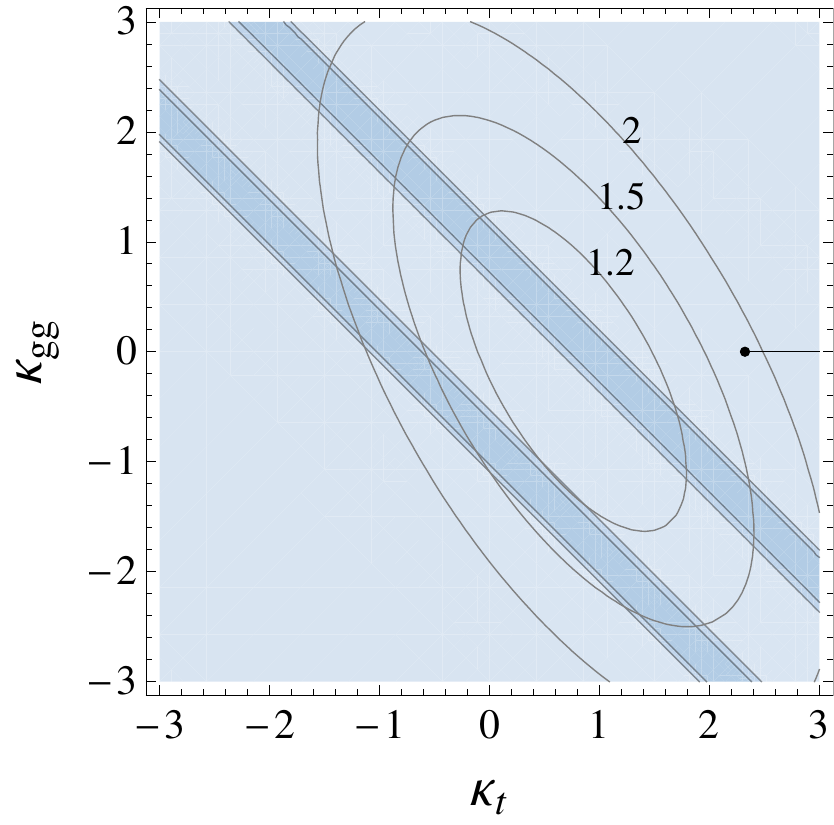}
\end{tabular}
 \end{center}
\caption{\label{fig:chi2}{\small Left: Dependence on the new-physics mass of the cross section of the $gg \to VV$ process in the off-shell region, normalized to the SM cross section for the same process. Right: Regions allowed at 1 and 2 $\sigma$ by the on-peak coupling constraints (diagonal bands), compared to the off-shell iso-cross section lines for the $gg \to VV$ process at 1.2, 1.50 and 2 times the SM cross section. The dot and black line correspond to the $\kappa_{gg}=0$ case where the CMS bound can be recast to the constraint $\kappa_t<2.3$.}}
\end{figure}

The form factor associated with $\kappa_{gg}$ depends on the mass of the particle(s) running in the loop, however this dependence is suppressed for large values of the mass, as the loop can be replaced by a higher order operator.
To quantify this statement, we considered the process $gg \to ZZ \to l^+ l^- {l'}^+ {l'}^-$ and study the dependence of the off-shell cross section. Following the experimental study \cite{Khachatryan:2014iha}, by off-shell cross section we define the cross section integrated over a range $\s>330$ GeV after the basic experimental cuts.
Note however that this is an intermediate definition between the off-shell region defined in \cite{Khachatryan:2014iha} ($m_{4\ell} > 220 $GeV) and the signal-enriched region ($m_{4\ell} > 330$ GeV plus a cut on a matrix element MELA discriminant), since our cross sections is estimated before the MELA discriminant cut. Our calculation has been performed by using the code \texttt{gg2VV} \cite{gg2vv}, modified by us in order to include the effect of a heavy particle loop. To be specific, all the results in this paper are based on a top-like fermion running in the loop.
In the left panel of Figure \ref{fig:chi2} we show the dependence of the off-shell cross section as a function of the new fermion mass, fixing $\kappa_t = 1$ (so that the top loop is standard) and $\kappa_{gg} = 1$.
From the plot we can see that the value starts becoming fairly independent on the mass between $500$ GeV and $1$ TeV: considering the typical direct bounds from LHC on coloured new particles, the approximation of infinite mass is very effective.
In the right panel of Figure \ref{fig:chi2}  we also show iso-cross section contours as a function of $\kappa_t$ and $\kappa_{gg}$ together with the region of parameters preferred by on-peak measurements and for large new physics mass.
The degenerate direction $\kappa_t + \kappa_{gg} \sim \pm 1$ is now resolved, thus an interpretation of the experimental bound in terms of these two parameters can be useful to resolve this degeneracy and directly probe new physics loops in the Higgs to gluon coupling.
In table 1 of \cite{Khachatryan:2014iha}, the off-shell signal-enriched region after MELA discrimination is shown to contain $11$ events against the expected $11.4 \pm 0.8$: these numbers, when compared to the $1.8$ expected $gg \to VV \to 4\ell$ events, show that at $95$\% C.L. the study is sensitive to an approximate doubling of the $gg \to VV$ cross section.
The actual bound is extracted from a signal enhanced region, thanks to a matrix element discriminant.
However, the experimental analysis is performed in absence of the new physics loop, therefore a direct reinterpretation of the bound cannot be done because the energy dependence and kinematics of the top and new physics loops can differ substantially. We leave a full analysis for a future publication, and for the experimental collaborations.

The main advantage of our proposal is that the analysis results will contain much more information that the interpretation in terms of the Higgs width.
The Higgs width bound can be recovered in the limit $\kappa_{gg} = 0$: in this case, a rescaling of the width by a factor $\Gamma_H = \xi\, \Gamma_H^{\rm SM}$ can be compensated on-peak by $\kappa_t = \sqrt{\xi}$.
Therefore, the published bound $\Gamma_H < 5.4\, \Gamma_H^{\rm SM}$ at 95\% CL can be re-expressed as an upper bound on $\kappa_t < \sqrt{5.4} = 2.3$ (dot in Figure \ref{fig:chi2}, with the excluded line on the right side). We can see that this dot is close to the iso-contour giving a doubled cross section, consistently with the previous remark about numbers of events.
This bound, however, only applies to positive values of $\kappa_t$ and it does not cover the possibility of negative couplings, for which the interference term has different sign.
To compare this bound with the constraints from the on-peak measurements, in Figure \ref{fig:chi2} we show the result of the fit \cite{flament} where we leave the two parameters $\kappa_t$ and $k_{gg}$ free, profile over the new physics loop in the di-photon coupling in order to fit the di-photon signal, and set all the other couplings to the standard model value.
The plot clearly show that the on-peak measurements have a degeneracy for $\kappa_t + \kappa_{gg} \sim \pm 1$, which is removed by the off-shell measurement.
An analysis of the off-shell measurement in terms of the two parameters would clearly add important information on the Higgs couplings, independently on its width.
It is finally interesting to quote the direct bound on $\kappa_t$ coming from the measurement of the $t \bar{t} h$ associated production. Assuming that the decay rate in $b \bar{b}$ is standard model like, the cross section is simply proportional to $\kappa_t^2$: from the CMS published bound \cite{Chatrchyan:2013yea} based on an integrated luminosity of 5 fb$^{-1}$ at 8 TeV, we can extract $|\kappa_t| < \sqrt{5.8} = 2.4$, while the ATLAS measurement with full 8 TeV dataset leads to $|\kappa_t| < \sqrt{4.1} = 2.02$ at 95\%CL \cite{ATLAS-CONF-2014-011}.\\

The simple analysis in this letter shows that the off-shell measurement of $gg \to ZZ$, performed by CMS \cite{Khachatryan:2014iha}, can be recast to a model-independent bound on the couplings of the Higgs to the top and on the contribution of new states in the Higgs to gluon coupling. In this perspective, the parametrisation of the Higgs couplings that we proposed in previous publications \cite{llodra,flament}, where two independent parameters are used for the tree level top coupling and new physics loops, are very useful for a model-independent interpretation of off-shell measurements.

\section*{Acknowledgements}
We thank S. Moretti for pointing out to us the CMS measurement, and initiating the discussion leading to this letter. We also 
thank S. Gascon-Shotkin for useful discussion. We acknowledge partial support from the Labex-LIO (Lyon Institute of Origins) 
under grant ANR-10-LABX-66 and FRAMA (FR3127, F\'ed\'eration de Recherche ``Andr\'e Marie Amp\`ere''). AD is partially supported 
by Institut Universitaire de France. 

\providecommand{\href}[2]{#2}\begingroup\raggedright\endgroup


\begin{thebibliography}{10}

\bibitem{atlas_higgs}
\textbf{ATLAS} Collaboration, "ATLAS-CONF-2014-009" ,"Updated coupling
  measurements of the Higgs boson with the ATLAS detector using up to 25
  fb$^{−1}$ of proton-proton collision data",
  http://cds.cern.ch/record/1670012.

\bibitem{cms_higgs}
{\bf CMS Collaboration} Collaboration, S.~Chatrchyan {\em et al.},
  ``{Observation of a new boson with mass near 125 GeV in pp collisions at
  $\sqrt{s}$ = 7 and 8 TeV},'' {\em JHEP} {\bf 1306} (2013) 081,
\href{http://www.arXiv.org/abs/1303.4571}{{\tt 1303.4571}}.

\bibitem{lhcxswg_reco}
{\bf LHC Higgs Cross Section Working Group} Collaboration, A.~David {\em et
  al.}, ``{LHC HXSWG interim recommendations to explore the coupling structure
  of a Higgs-like particle},''
\href{http://www.arXiv.org/abs/1209.0040}{{\tt 1209.0040}}.

\bibitem{passarino}
N.~Kauer and G.~Passarino, ``{Inadequacy of zero-width approximation for a
  light Higgs boson signal},'' {\em JHEP} {\bf 1208} (2012) 116,
\href{http://www.arXiv.org/abs/1206.4803}{{\tt 1206.4803}}.

\bibitem{melnikov}
F.~Caola and K.~Melnikov, ``{Constraining the Higgs boson width with ZZ
  production at the LHC},'' {\em Phys.Rev.} {\bf D88} (2013) 054024,
\href{http://www.arXiv.org/abs/1307.4935}{{\tt 1307.4935}}.

\bibitem{campbell}
J.~M. Campbell, R.~K. Ellis, and C.~Williams, ``{Bounding the Higgs width at
  the LHC using full analytic results for $gg \to e^- e^+ \mu^- \mu^+$},'' {\em
  JHEP} {\bf 1404} (2014) 060,
\href{http://www.arXiv.org/abs/1311.3589}{{\tt 1311.3589}}.

\bibitem{Khachatryan:2014iha}
{\bf CMS Collaboration} Collaboration, V.~Khachatryan {\em et al.},
  ``{Constraints on the Higgs boson width from off-shell production and decay
  to $Z$-boson pairs},''
\href{http://www.arXiv.org/abs/1405.3455}{{\tt 1405.3455}}.

\bibitem{englert}
C.~Englert and M.~Spannowsky, ``{Limitations and Opportunities of Off-Shell
  Coupling Measurements},''
\href{http://www.arXiv.org/abs/1405.0285}{{\tt 1405.0285}}.

\bibitem{llodra}
G.~Cacciapaglia, A.~Deandrea, and J.~Llodra-Perez, ``{Higgs $\to \gamma \gamma$
  beyond the Standard Model},'' {\em JHEP} {\bf 0906} (2009) 054,
\href{http://www.arXiv.org/abs/0901.0927}{{\tt 0901.0927}}.

\bibitem{flament}
G.~Cacciapaglia, A.~Deandrea, G.~D. La~Rochelle, and J.-B. Flament, ``{Higgs
  couplings beyond the Standard Model},'' {\em JHEP} {\bf 1303} (2013) 029,
\href{http://www.arXiv.org/abs/1210.8120}{{\tt 1210.8120}}.

\bibitem{gg2vv}
N.~Kauer, ``{Interference effects for H $\to$ WW/ZZ $\to
  \ell\bar{\nu}_\ell\bar{\ell}\nu_\ell$ searches in gluon fusion at the LHC},''
  {\em JHEP} {\bf 1312} (2013) 082,
\href{http://www.arXiv.org/abs/1310.7011}{{\tt 1310.7011}}.

\bibitem{Chatrchyan:2013yea}
{\bf CMS Collaboration} Collaboration, S.~Chatrchyan {\em et al.}, ``{Search
  for the standard model Higgs boson produced in association with a top-quark
  pair in pp collisions at the LHC},'' {\em JHEP} {\bf 1305} (2013) 145,
\href{http://www.arXiv.org/abs/1303.0763}{{\tt 1303.0763}}.

\bibitem{ATLAS-CONF-2014-011}
\textbf{ATLAS} Collaboration, "ATLAS-CONF-2014-011" ,"Search for the Standard
  Model Higgs boson produced in association with top quarks and decaying to
  b-bbar in pp collisions at sqrt(s)= 8 TeV with the ATLAS detector at the
  LHC", http://cds.cern.ch/record/1670532.

\end{thebibliography}
\end{document}